\documentclass[3p,times,procedia]{elsarticle}
\usepackage{nupha_ecrc}

\volume{00}

\firstpage{1}

\journalname{Nuclear Physics A}

\runauth{}

\jid{nupha}

\jnltitlelogo{Nuclear Physics A}

\usepackage{amssymb}

\usepackage[figuresright]{rotating}

\begin{document}

\begin{frontmatter}




\title{Neutral meson production and correlation with charged hadrons in pp and Pb-Pb collisions with the ALICE experiment at the LHC}


\author{Astrid Vauthier \\on behalf of the ALICE Collaboration}

\address{LPSC, Universit\'e Grenoble-Alpes, CNRS/IN2P3 - 53, Avenue des Martyrs 38026 Grenoble Cedex}

\begin{abstract}
Among the probes used to investigate the properties of the Quark-Gluon Plasma, the measurement of the energy loss of high-energy partons can be used to put constraints on energy-loss models and to ultimately access medium characteristics, such as the energy density or the temperature. The study of two-particle correlations allows us to obtain very different constraints compared to the nuclear modification factor. In particular, the correlation of charged hadrons with high energy $\pi^{0}$ or direct photons is believed to give a measurement of the parton energy loss and insights into the medium-induced modification of the fragmentation process. 

High energy neutral pions are reconstructed using the ALICE electromagnetic calorimeters EMCal and PHOS, and the charged particles are detected by the main tracking detectors ITS and TPC.
 
In these proceedings, the measurement of neutral mesons at $\sqrt{s} = $2.76 TeV in pp collisions are presented, as well as the measurements of azimuthal $\pi^{0}$-hadron correlations in pp and Pb-Pb collisions at $\sqrt{s_{\rm{NN}}} = 2.76$~TeV, and the extracted per-trigger yield modification factor ($I_{AA}$). Comparisons with theoretical model calculations are also added.
\end{abstract}


\end{frontmatter}


\section{Introduction}

During heavy-ion collisions, a hot, dense, deconfined and strongly interacting QCD medium is formed: the so called quark-gluon plasma. Studying several collision systems aims both at probing such a medium and understanding the strong interaction.
In heavy-ion collisions, the medium properties and the in-medium modifications can be measured, while pp collisions are used as a reference, as a test of perturbative QCD (pQCD), and as constraints to the parton distribution (PDFs) and the fragmentation (FF) functions. p-A collisions are used to discriminate cold and hot nuclear matter effects but will not be discussed here.
The measurement of neutral meson production allows us to obtain a rough approach to parton energy loss through its nuclear modification factor $R_{\rm{AA}}$. 
The study of $\pi^0$-hadron correlations permits us to obtain other constraints to parton energy loss and is a necessary step to access photon-hadron correlations that provide a calibration of the parton energy scale for energy loss studies.

%
%
%

\section{Neutral meson identification \label{sec:NeutralMesonID}}

The photons coming from neutral meson decays are reconstructed using the two ALICE electromagnetic calorimeters, PHOS and EMCal (homogeneous and sampling calorimeters respectively), but also with the photon conversion method, PCM, that reconstructs photons that convert in the central tracking detectors.

Two identification methods are used with the calorimeters. The first one consists in combining pairs of clusters to reconstruct the invariant mass of the two decay photons coming from the neutral mesons. The candidates whose invariant mass falls close to the $\pi^0$ or $\eta$ mass are kept and the combinatorial background is treated statistically, as shown in Fig. \ref{Fig:MesonID}, top panel. 

The second analysis uses elongated clusters made of the merged electromagnetic showers of both decay photons. The elongation can be quantified by the largest squared eigenvalue $\sigma_{\rm{long}}^2$ (see Fig. \ref{Fig:MesonID}, bottom panel) of the cluster's energy decomposition in the (pseudorapidity, azimuth) plane of EMCal, which, due to the decay angle variation decreases with increasing meson $p_{\rm{T}}$. $\pi^0$ can then be selected with an energy dependent cut on $\sigma_{\rm{long}}^2$. This method will be referred as "single cluster analysis" in the following. 

%
%
%

\section{Measurement of neutral meson production}

\begin{figure}[!htbp]
\begin{minipage}[c]{.3\linewidth}
	\centering\includegraphics[width = \linewidth]{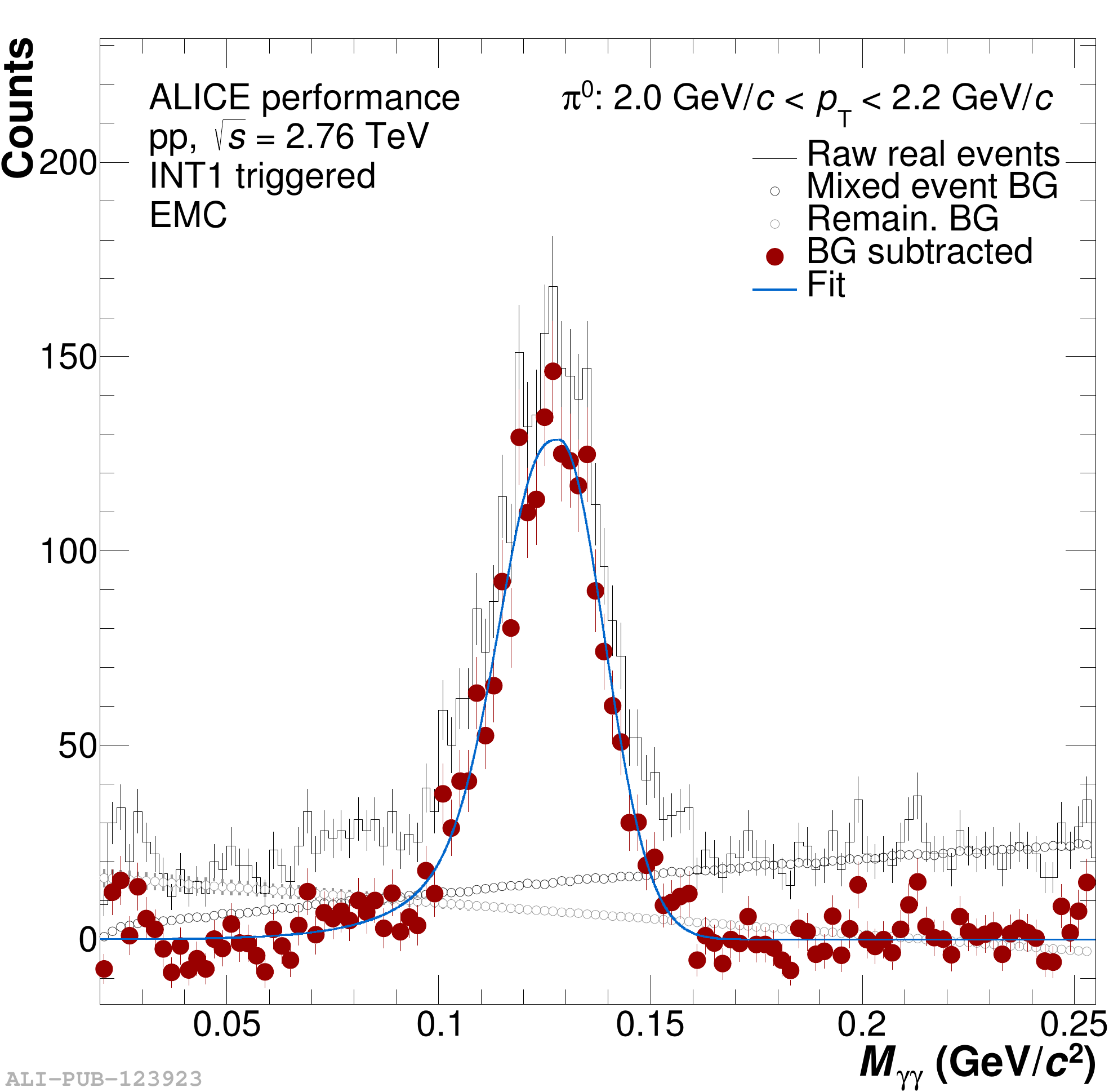}
	
	\includegraphics[width = \linewidth]{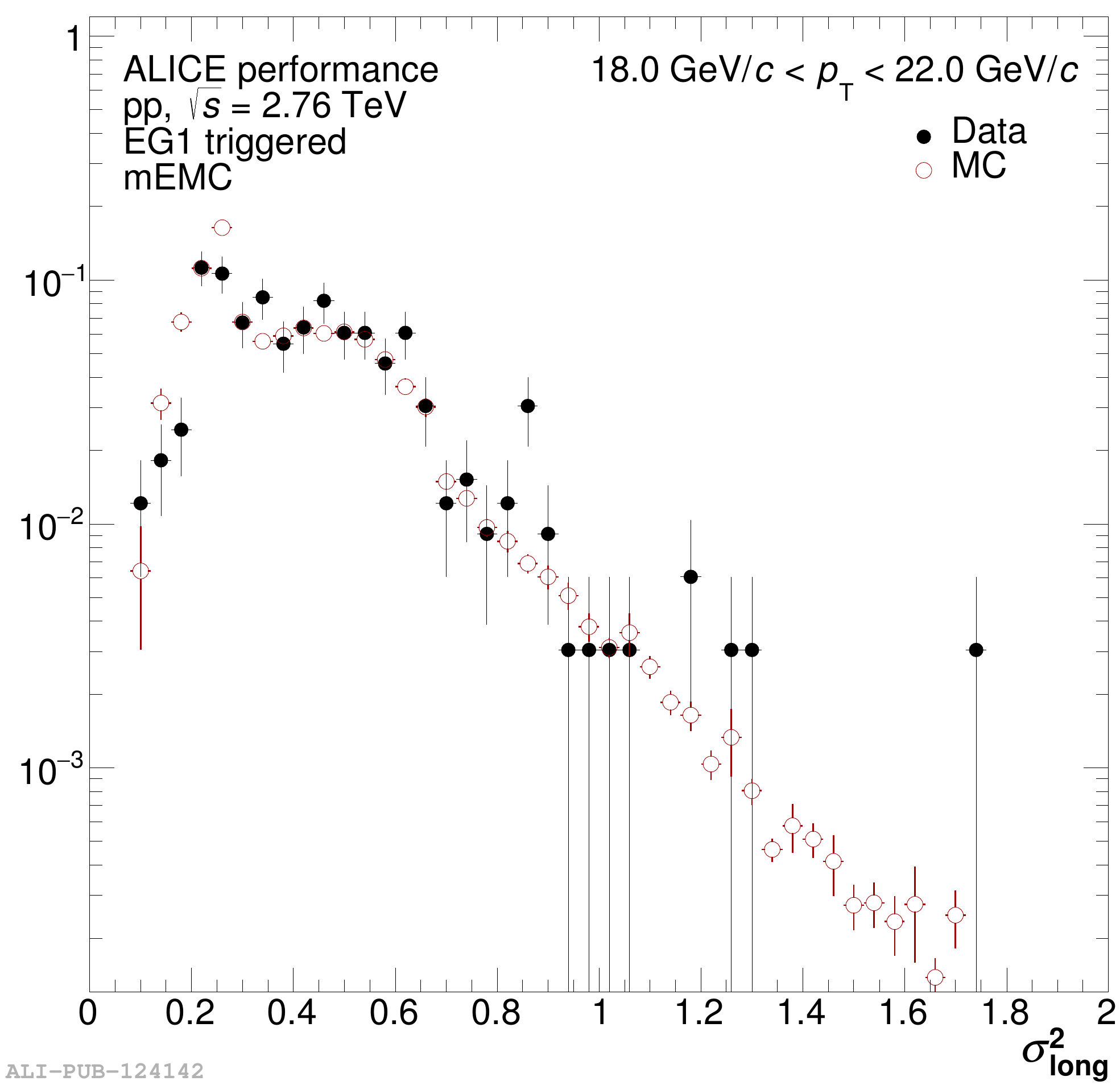}
	
	\caption{Top: $\gamma\gamma$ invariant mass distribution with EMCal in pp collisions. Bottom: $\sigma_{\rm{long}}^2$ distribution in pp collisions.\label{Fig:MesonID}}
\end{minipage}
\hspace{.2cm}
\begin{minipage}[c]{.68\linewidth}
	\centering\includegraphics[width = .49\linewidth]{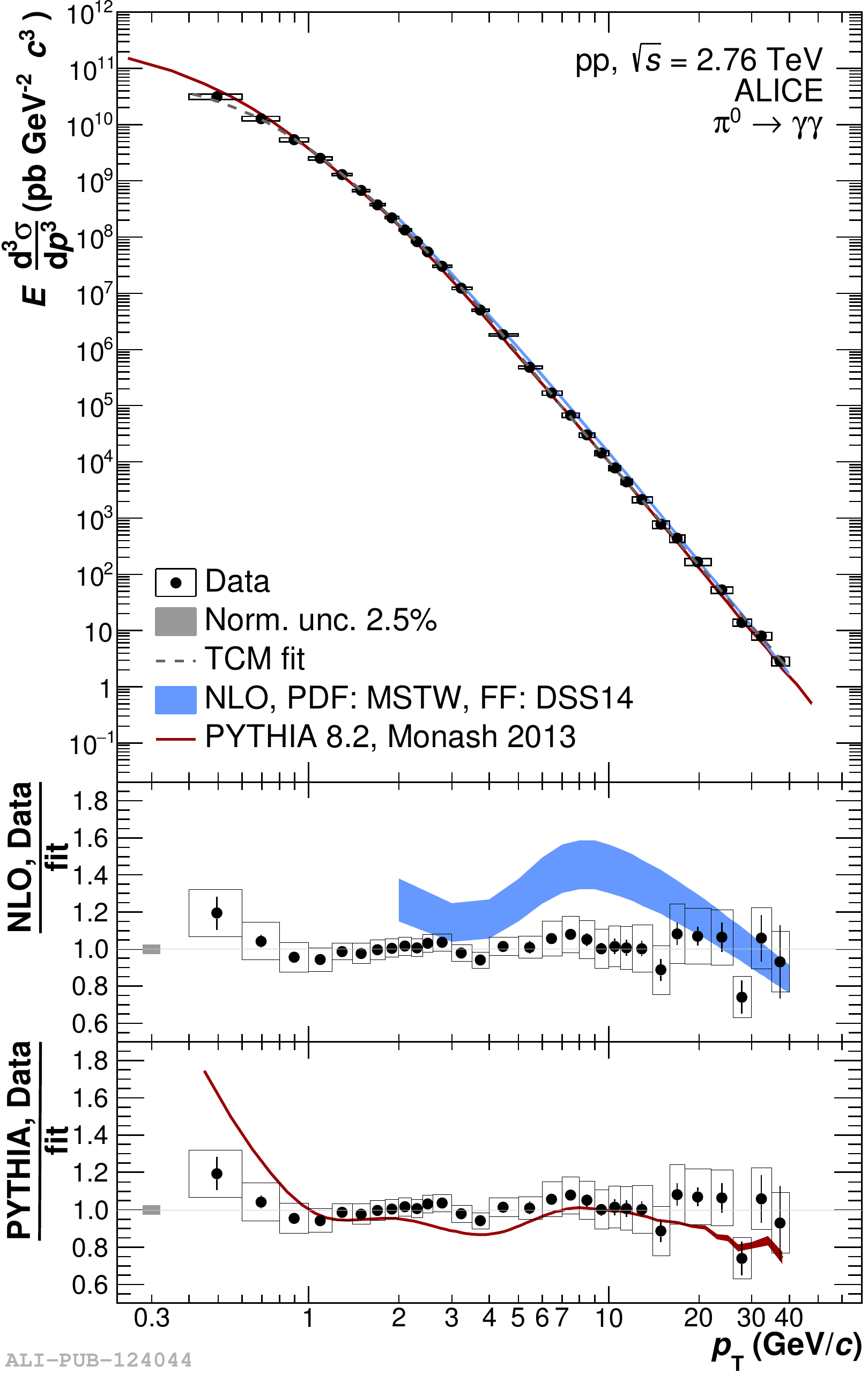}
	\centering\includegraphics[width = .49\linewidth]{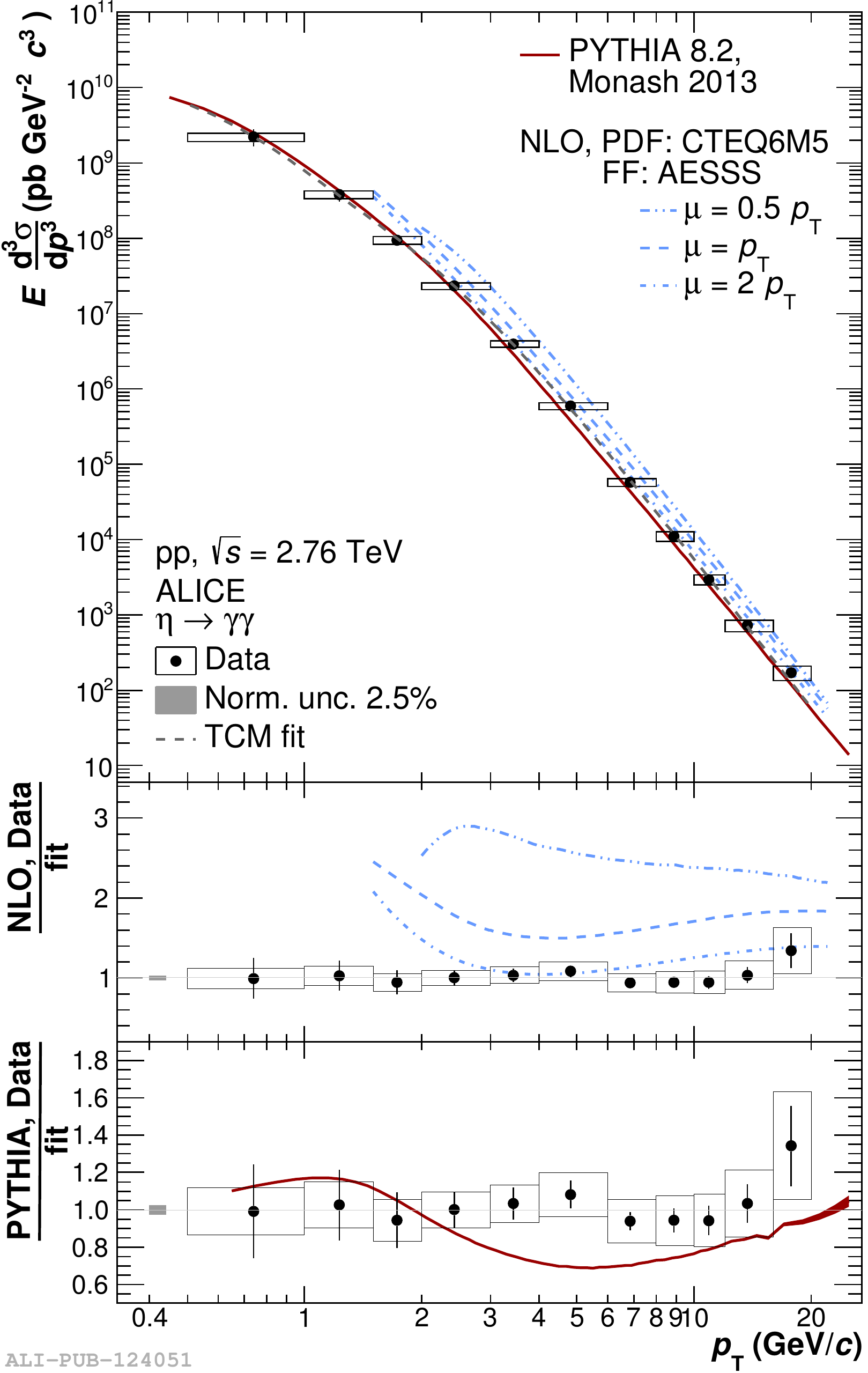}
	\caption{Left: $\pi^{0}$ invariant cross section in pp collisions at $\sqrt{s} = 2.76$ TeV, obtained from PCM, PHOS and EMCal combined measurements~\cite{NeutralMesonPaper}. The results are compared to NLO pQCD (MSTW + DSS14~\cite{DSS14}) and PYTHIA calculations. Right: $\eta$ invariant cross section in pp collisions at $\sqrt{s} = 2.76$ TeV, obtained from PCM and EMCal combined measurements~\cite{NeutralMesonPaper}. The results are compared to NLO pQCD (CTEQ + AES~\cite{AES}) and PYTHIA calculations.	\label{Fig:XSecpp}}
	
\end{minipage}
\end{figure}

The cross section for neutral pions and $\eta$ mesons is shown in the left and right panels of the Fig. \ref{Fig:XSecpp} for pp collisions at $\sqrt{s} = 2.76$~TeV~\cite{NeutralMesonPaper}. The cross section for $\pi^0$ results from a combination of PCM, PHOS and EMCal measurements from 2011~\cite{NeuralMesonPbPb} and 2013 data while the $\eta$ cross section results from a combination of PCM and EMCal measurements. The new EMCal measurement using the single cluster analysis allows us to extend the $p_{\rm{T}}$ reach from 20 GeV/$c$ up to 40 GeV/$c$ for $\pi^0$, and the new measurement allows us to reach a $p_{\rm{T}}$ up to 20 GeV/$c$ for the $\eta$. 

The results are compared to PYTHIA 8.2 Monash 2013, which agrees better with the data for $\pi^0$ than for the $\eta$. Comparisons with NLO pQCD calculations lead to the same disagreement for the $\eta$ than with the $\pi^0$~\cite{DisagreeCTEQ} when CTEQ PDFs and AES~\cite{AES} for the fragmentation function are used, but the discrepancy is much reduced for $\pi^0$ when MSTW PDFs with DSS14~\cite{DSS14} for the fragmentation function are used.
Combined with the Pb-Pb measurement at $\sqrt{s_{\rm{NN}}} = $ 2.76 TeV~\cite{NeuralMesonPbPb}, these results will allow us to obtain the nuclear modification factor, $R_{\rm{AA}}$, for the $\eta$ meson as well as to extend the $\pi^0$ $R_{\rm{AA}}$ measurement~\cite{RAAPi0} up to 20 GeV/$c$. 

Yet differential measurements such as $\pi^0$-hadron correlations are also needed to have a better understanding of the in-medium energy loss. 

\section{Measurement of $\pi^0$-hadron correlations}

The energy loss of hard partons in the deconfined medium can be probed by looking at the charged hadrons correlated with a high momentum $\pi^0$, that are thought to tag hard QCD processes.

The analysis~\cite{pi0HCorr} consists in associating high momentum $\pi^0$ with charged hadrons in order to study the angular correlation between the charged hadrons and the "trigger" $\pi^0$, $\Delta\varphi = \varphi^{\pi^{0}} - \varphi^{hadron}$ where $\varphi$ is the azimuthal position of the particle. The trigger $\pi^0$ are identified using the single cluster analysis described in Section \ref{sec:NeutralMesonID} and to reject the hadrons which don't come from the fragmentation, the underlying event is subtracted with the ZYAM method, using a flat background in pp collisions, and a background estimated with the measured anisotropic flow $v_{n}$ for Pb-Pb collisions~\cite{FlowTrigg,FlowHadrons} and given by the equation $B(\Delta\varphi) =  B_{0}\left(1+2\sum_{n =2}^5 v_{n}\cos(n\Delta\varphi)\right)$.

\begin{figure}[!htbp]
	\centering\includegraphics[width = .49\linewidth]{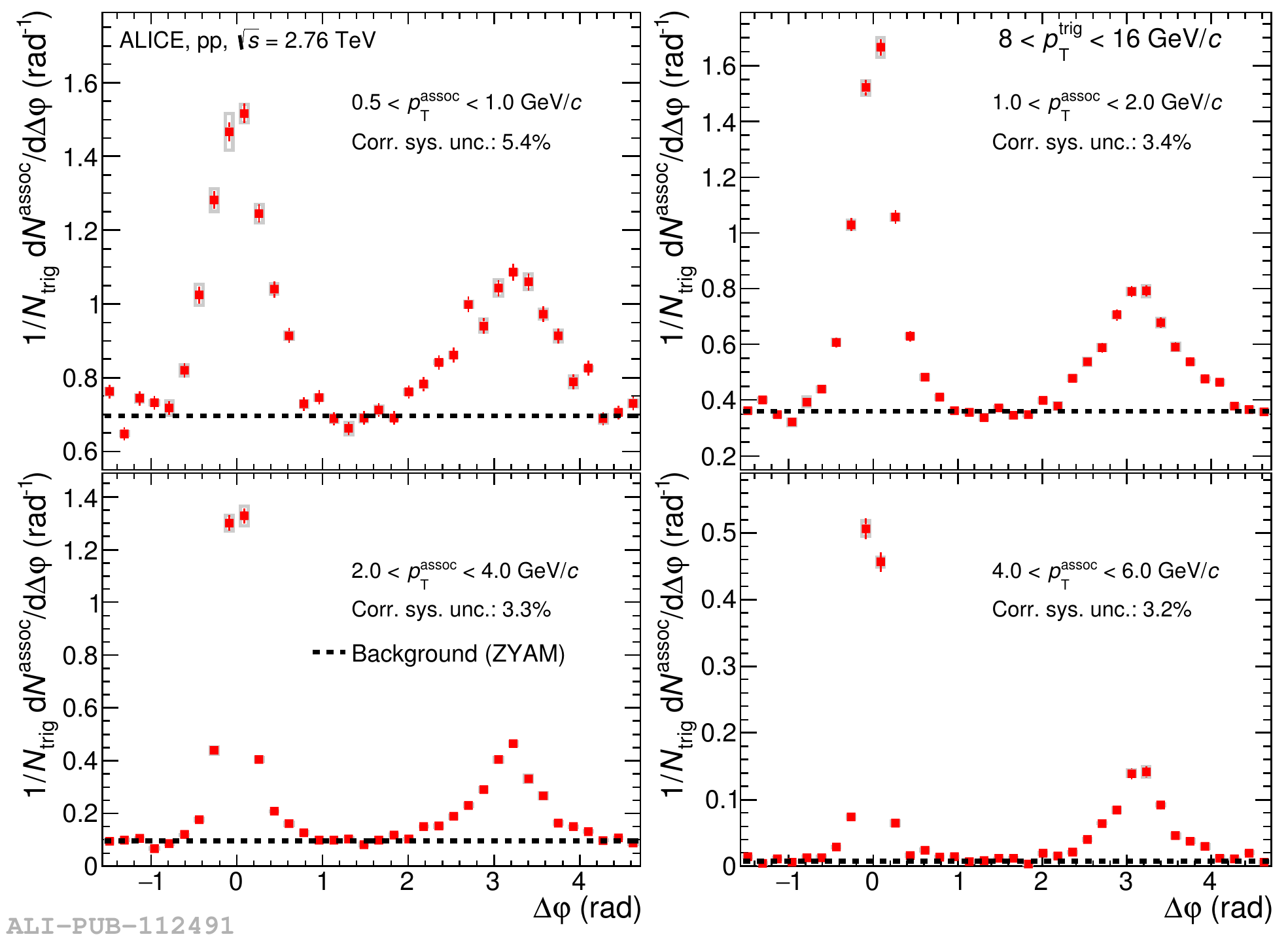}
	\hspace{.1cm}
	\includegraphics[width = .49\linewidth]{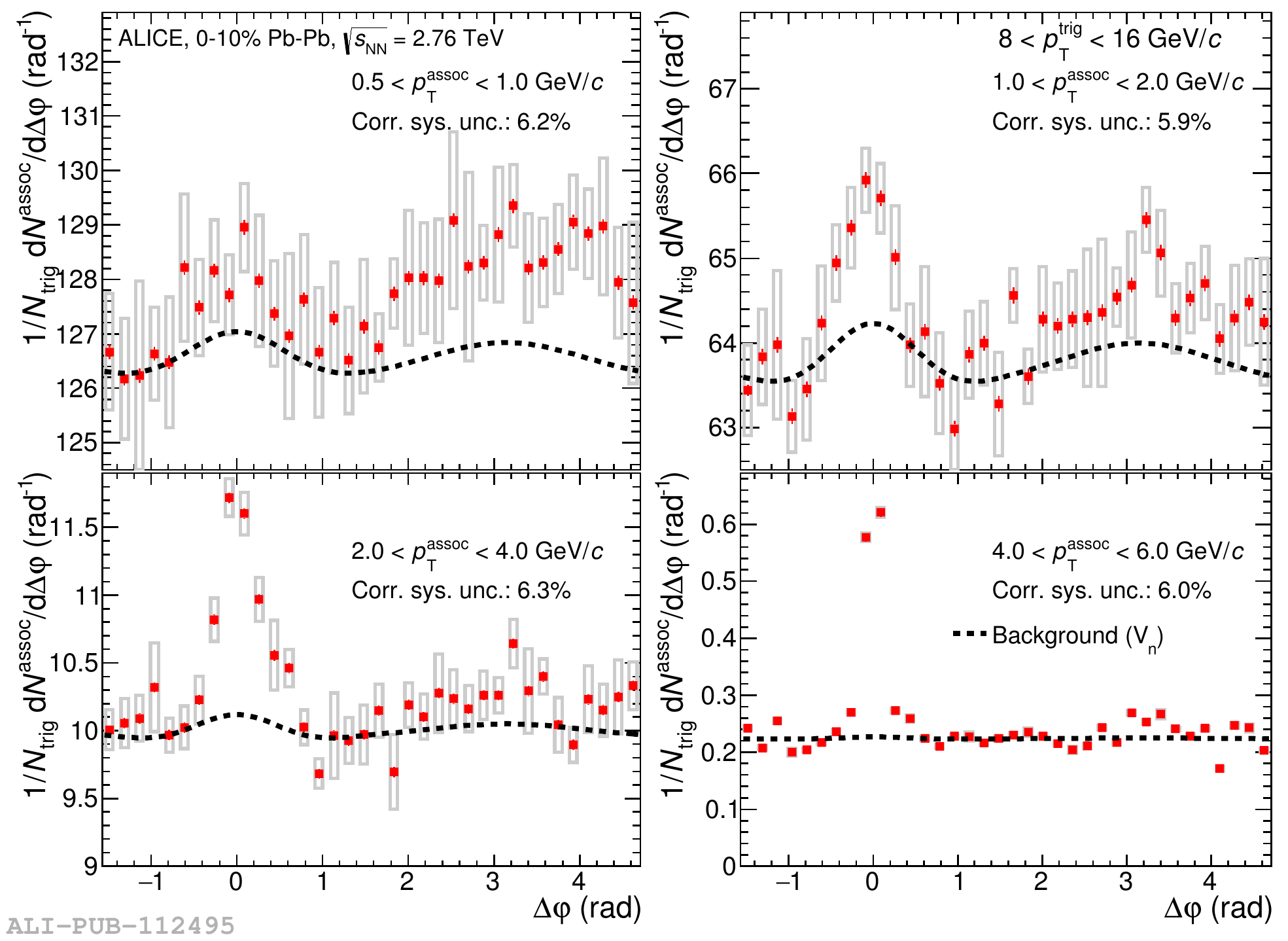}
	\caption{The angular correlation, $C(\Delta\varphi)$, for a trigger $\pi^0$ between 8 and 16 GeV/$c$ and 4 chosen $p_{\rm{T}}^{assoc}$ bins for the associated charged hadrons for pp (left) and Pb-Pb (right) collisions~\cite{pi0HCorr}.}
	\label{Fig:PerTriggerYield}
\end{figure}

The angular correlations $C(\Delta\varphi)$, are shown in Fig.~\ref{Fig:PerTriggerYield} for a trigger $\pi^0$ between 8 and 16 GeV/$c$ and four chosen $p_{\rm{T}}$ bins for the associated charged hadrons, for pp collisions on the left and the most central Pb-Pb collisions on the right. The red points represent the total correlation, and the dashed line the background contribution. The per-trigger yield $J(p_{\rm{T}}^{\pi^{0}},~p_{\rm{T}}^{h^{\pm}})$ is calculated by integrating $C(\Delta\varphi) - B(\Delta\varphi)$ over the selected $\Delta\varphi$ range.

The medium-induced per-trigger yield modification factor $I_{\rm{AA}}(p_{\rm{T}}^{\pi^{0}},~p_{\rm{T}}^{h^{\pm}}) = \frac{J_{\rm{AA}}(p_{\rm{T}}^{\pi^{0}},~p_{\rm{T}}^{h^{\pm}})}{J_{\rm{pp}}(p_{\rm{T}}^{\pi^{0}},~p_{\rm{T}}^{h^{\pm}})}$ is studied in two regions: the "near side" and "away side" which are found respectively around and opposite to the trigger $\pi^0$. 

\begin{figure}[!htbp]
	\centering\includegraphics[width = .49\linewidth]{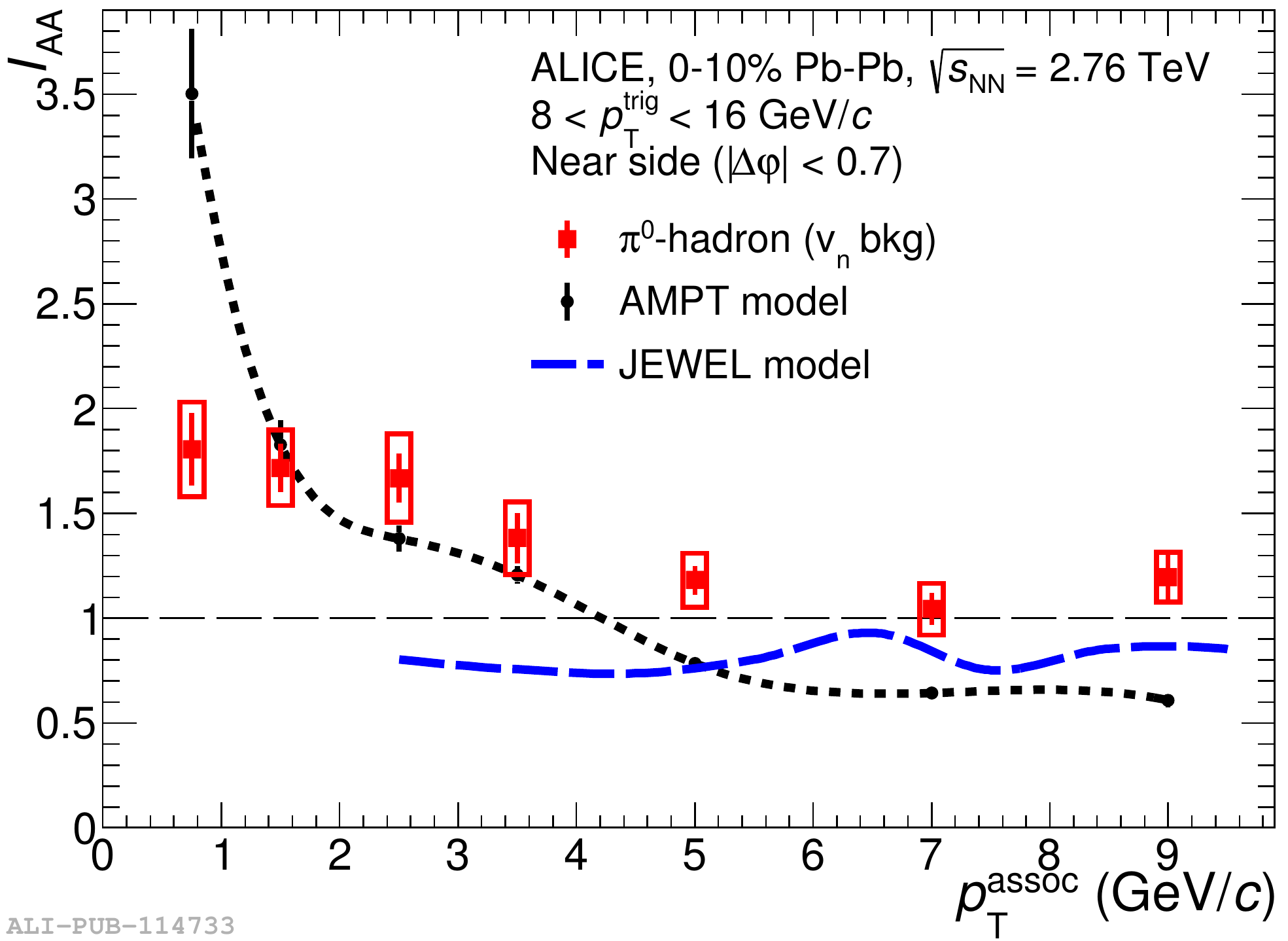}
	\hspace{.1cm}
	\includegraphics[width = .49\linewidth]{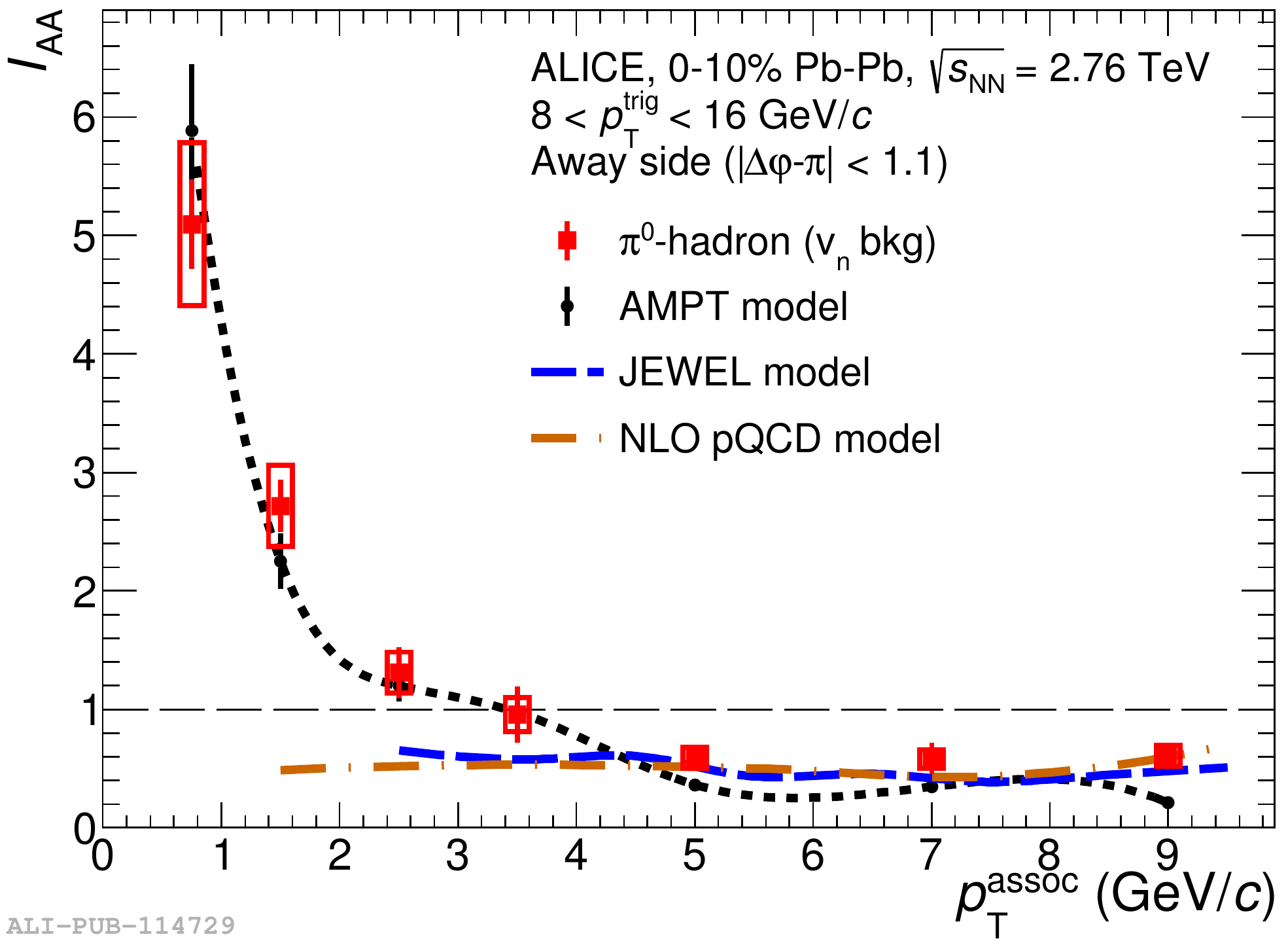}
	\caption{The medium induced per-trigger yield modification factor $I_{\rm{AA}}$ for near side (left) and away side (right)~\cite{pi0HCorr}.}
	\label{Fig:IAA}
\end{figure}

On the near side (Fig.~\ref{Fig:IAA}, left), an enhancement is observed at low $p_{\rm{T}}$, maybe due to a modification of the fragmentation function and/or of the quark to gluon jet ratio. Such a behavior could also be caused by a harder parton $p_{\rm{T}}$ spectrum. On the away side (Fig.~\ref{Fig:IAA}, right), a suppression attributed to parton energy loss is observed at high $p_{\rm{T}}$ in Pb-Pb collisions. At low $p_{\rm{T}}$, a large enhancement is observed, which was unexpected but could be explained by $k_{\rm{T}}$ broadening, medium excitation or by the presence of fragments from radiated gluons, all those phenomena being related to the jet interactions with the medium. Both sides show a good agreement with a previous measurement of di-hadron correlations~\cite{DiHCorr}.

This suppression at high $p_{\rm{T}}$ was already observed at RHIC energies ~\cite{pi0HCorrPHENIX}, but the low $p_{\rm{T}}$ enhancement is larger in the measurement by ALICE. As the higher harmonics have only been subtracted in the ALICE analysis, results cannot be directly compared.

These results can be compared with theoretical models. On the near side, the enhancement is reproduced only by the AMPT model~\cite{AMPT}, except below 1 GeV/$c$ where it overestimates the enhancement. On the away side, the high $p_{\rm{T}}$ suppression is qualitatively described by all the models but only AMPT decribes the low $p_{\rm{T}}$ enhancement. In this model, the low $p_{\rm{T}}$ enhancement is a consequence of the increase of soft particles as a result of the interactions between the jet and the QCD medium.

\section{Conclusions and outlook}

The new measurements of the $\pi^0$ and $\eta$ energy $p_{\rm{T}}$ spectra have been compared to PYTHIA and NLO pQCD calculations and will allow us to calculate the $\eta$ $R_{\rm{AA}}$ and to extend the published $\pi^0$ nuclear modification factor to high $p_{\rm{T}}$. The medium-induced per-trigger yield modification factor from $\pi^0$-hadron correlations has also been presented for the most central Pb-Pb events both on the near and away sides. A suppression is observed on the away side which is attributed to parton energy loss and well described by all the models. Both sides show an enhancement at low $p_{\rm{T}}$, that can be quantitatively described only by the AMPT model. 
A next step is to access the medium induced parton energy loss with direct $\gamma$-hadron correlation. Results from STAR~\cite{GHCorrSTAR} and PHENIX~\cite{GHCorrPHENIX} already show a clear evidence of parton energy loss. The analysis is ongoing in ALICE via two methods: isolation~\cite{GHCorrNico} and statistical subtraction both of which use the $\pi^0$-hadron correlations result to subtract the background.





\bibliographystyle{elsarticle-num}
\bibliography{QMProceeding_AVauthier}

\begin{thebibliography}{10}
\expandafter\ifx\csname url\endcsname\relax
  \def\url#1{\texttt{#1}}\fi
\expandafter\ifx\csname urlprefix\endcsname\relax\def\urlprefix{URL }\fi
\expandafter\ifx\csname href\endcsname\relax
  \def\href#1#2{#2} \def\path#1{#1}\fi

\bibitem{NeutralMesonPaper}
S.~Acharya, et~al., {}\href {http://arxiv.org/abs/1702.00917}
  {\path{arXiv:1702.00917}}.

\bibitem{NeuralMesonPbPb}
A.~Morreale, Nuclear Physics A 956 (2016) 645 -- 648.

\bibitem{DSS14}
D.~de~Florian, R.~Sassot, M.~Epele, R.~J. Hern\'andez-Pinto, M.~Stratmann,
  Phys. Rev. D 91 (2015) 014035.

\bibitem{AES}
C.~A. Aidala, F.~Ellinghaus, R.~Sassot, J.~P. Seele, M.~Stratmann, Phys. Rev. D
  83 (2011) 034002.

\bibitem{DisagreeCTEQ}
B.~Abelev, et~al., Phys. Lett. B717 (2012) 162--172.

\bibitem{RAAPi0}
B.~Abelev, et~al., Eur. Phys. J. C74~(10) (2014) 3108.

\bibitem{pi0HCorr}
J.~Adam, et~al., Phys. Lett. B763 (2016) 238--250.

\bibitem{FlowTrigg}
B.~Abelev, et~al., Phys. Lett. B719 (2013) 18--28.

\bibitem{FlowHadrons}
K.~Aamodt, et~al., Phys. Lett. B708 (2012) 249--264.

\bibitem{DiHCorr}
K.~Aamodt, et~al., Phys. Rev. Lett. 108 (2012) 092301.

\bibitem{pi0HCorrPHENIX}
A.~Adare, et~al., Phys. Rev. Lett. 104 (2010) 252301.

\bibitem{AMPT}
G.-L. Ma, X.-N. Wang, Phys. Rev. Lett. 106 (2011) 162301.

\bibitem{GHCorrSTAR}
L.~Adamczyk, et~al., Physics Letters B 760 (2016) 689 -- 696.

\bibitem{GHCorrPHENIX}
A.~Adare, et~al., Phys. Rev. Lett. 111 (2013) 032301.

\bibitem{GHCorrNico}
N.~Arbor, Nuclear Physics A 904 (2013) 697c -- 700c.

\end{thebibliography}







\end{document}